\documentclass[conference]{IEEEtran}
\IEEEoverridecommandlockouts
\usepackage{cite}
\usepackage{amsmath,amssymb,amsfonts}
\usepackage{algorithmic}
\usepackage{graphicx}
\usepackage{textcomp}
\usepackage{xcolor}
\usepackage{caption}

\usepackage{placeins}
\usepackage{float}
\usepackage{cleveref}
\usepackage{xcolor}
\usepackage{listings}
 \usepackage{wrapfig}

\lstdefinestyle{yaml}{
     basicstyle=\color{blue}\fontsize{6pt}{6pt}\selectfont,
     rulecolor=\color{black},
     string=[s]{'}{'},
     stringstyle=\color{blue},
     comment=[l]{:},
     commentstyle=\color{black},
     morecomment=[l]{-}
 }

\usepackage{booktabs} 
\usepackage{enumitem}
\usepackage[table]{xcolor}
\usepackage{array}
\usepackage{subcaption}
\usepackage{tcolorbox}
\definecolor{tealcolor}{RGB}{0, 150, 136}
\definecolor{slateblue}{RGB}{0, 95, 90}
\newcommand{\hide}[1]{Redacted for review}

\def\BibTeX{{\rm B\kern-.05em{\sc i\kern-.025em b}\kern-.08em
    T\kern-.1667em\lower.7ex\hbox{E}\kern-.125emX}}
\makeatletter 
\newcommand{\linebreakand}{%
  \end{@IEEEauthorhalign}
  \hfill\mbox{}\par
  \mbox{}\hfill\begin{@IEEEauthorhalign}
}
\makeatother 

\begin{document}
\title{Enabling Reuse for Data-Sharing Pipelines in Federated Environments}

\author{
\IEEEauthorblockN{Sepideh Masoudi}
\IEEEauthorblockA{\textit{Information Systems Engineering}\\
\textit{TU Berlin}\\
Berlin, Germany\\
smi@ise.tu-berlin.de} 
\and
\IEEEauthorblockN{Maria C. Borges}
\IEEEauthorblockA{\textit{Information Systems Engineering}\\
\textit{TU Berlin}\\
Berlin, Germany\\
mb@ise.tu-berlin.de} 
\and
\IEEEauthorblockN{Eduardo Brito}
\IEEEauthorblockA{
\textit{Cybernetica AS} \&\\
\textit{University of Tartu}\\
Tartu, Estonia\\
eduardo.brito@cyber.ee\\
}
\linebreakand
\IEEEauthorblockN{Sebastian Werner}
\IEEEauthorblockA{\textit{Information Systems Engineering}\\
\textit{TU Berlin}\\
Berlin, Germany\\
sw@ise.tu-berlin.de} 
\and
\IEEEauthorblockN{Stefan Tai}
\IEEEauthorblockA{\textit{Information Systems Engineering}\\
\textit{TU Berlin}\\
Berlin, Germany\\
st@ise.tu-berlin.de}
}
\maketitle

\begin{abstract}
Data mesh architectures enable decentralized data sharing through domain-owned data products, but supporting diverse consumers in federated settings often requires customized data-sharing pipelines. As the number of consumers grows, this leads to a proliferation of pipelines, increasing design and maintenance complexity. We observe that such pipelines frequently exhibit substantial structural overlap. In this paper, we argue that reuse should serve as a guiding principle to address this challenge. We define reuse in data-sharing pipelines as the systematic use of existing data assets and transformation logic across pipelines, and identify reuse opportunities at both design time and runtime. We analyze the associated challenges and outline a reuse-oriented design approach, supported by a reference architecture. A preliminary evaluation demonstrates the potential of reuse to reduce redundancy and improve manageability, providing a pathway toward more scalable and sustainable federated data sharing.
\end{abstract}


\section{Introduction}\label{sec:introduction}As organizations generate ever-growing volumes of data, the range of use cases for this data also expands. To accommodate this diversity, data mesh architectures have recently been proposed~\cite{machado2022,dehghani2022}. Rather than funneling all data through a central team, the data mesh paradigm treats data as a business product, managed by the domain teams closest to its source. 
Each team prepares and serves analytical data to different consumers as a \emph{data product}, an independently deployable component encompassing data, metadata, transformation code, and the required infrastructure~\cite{driessen2025,goedegebuure2024data}.

While data meshes were initially proposed for intra-organizational use, they also extend naturally to federated data-sharing contexts~\cite{sedlak2024}. Cross-organizational data sharing is becoming increasingly important, and the well-defined interfaces and metadata of data products allow organizations to easily discover and derive value from each other's data. However, this setting also introduces new challenges related to control. A shared data product must satisfy the requirements of both providers and consumers, including constraints on permitted processing purposes, data quality guarantees, and compliance obligations~\cite{falconi2024}. 
\emph{Data-sharing pipelines}, which are sequences of transformation stages that manipulate data to satisfy such requirements, provide a technical mechanism to enforce these constraints in federated data products~\cite{werner_teadal_2023}.

As federated data sharing scales, accommodating the growing diversity of consumer requirements becomes increasingly complex. The same underlying data often needs to be assembled, filtered, or structured differently depending on the consuming organization. A straightforward approach is to construct a dedicated data pipeline for each consumer. However, this leads to a proliferation of pipelines that is difficult to manage, increases operational overhead, and limits scalability~\cite{driessen2025,dehghani2022}.

A key observation is that these pipelines are rarely independent. Instead, they often exhibit substantial structural overlap, for example, in shared transformation steps such as anonymization, filtering, or formatting~\cite{werner_teadal_2023}. This suggests an opportunity for reuse, understood as the systematic use of existing data assets and transformation logic across multiple pipelines to avoid redundant implementation and improve consistency.

We therefore argue that reuse should serve as a guiding principle in the design of federated data-sharing environments. In particular, reuse can be realized at different stages of the system lifecycle. At design time, reusable transformation components and pipeline templates can be defined and composed across multiple pipelines. At runtime, shared data products can be accessed by multiple consumers through well-defined interfaces. By leveraging such reuse opportunities, providers can reduce duplication, lower operational costs, and simplify maintenance.

\begin{figure*}[t]
\centering
\includegraphics[width=0.8\textwidth]{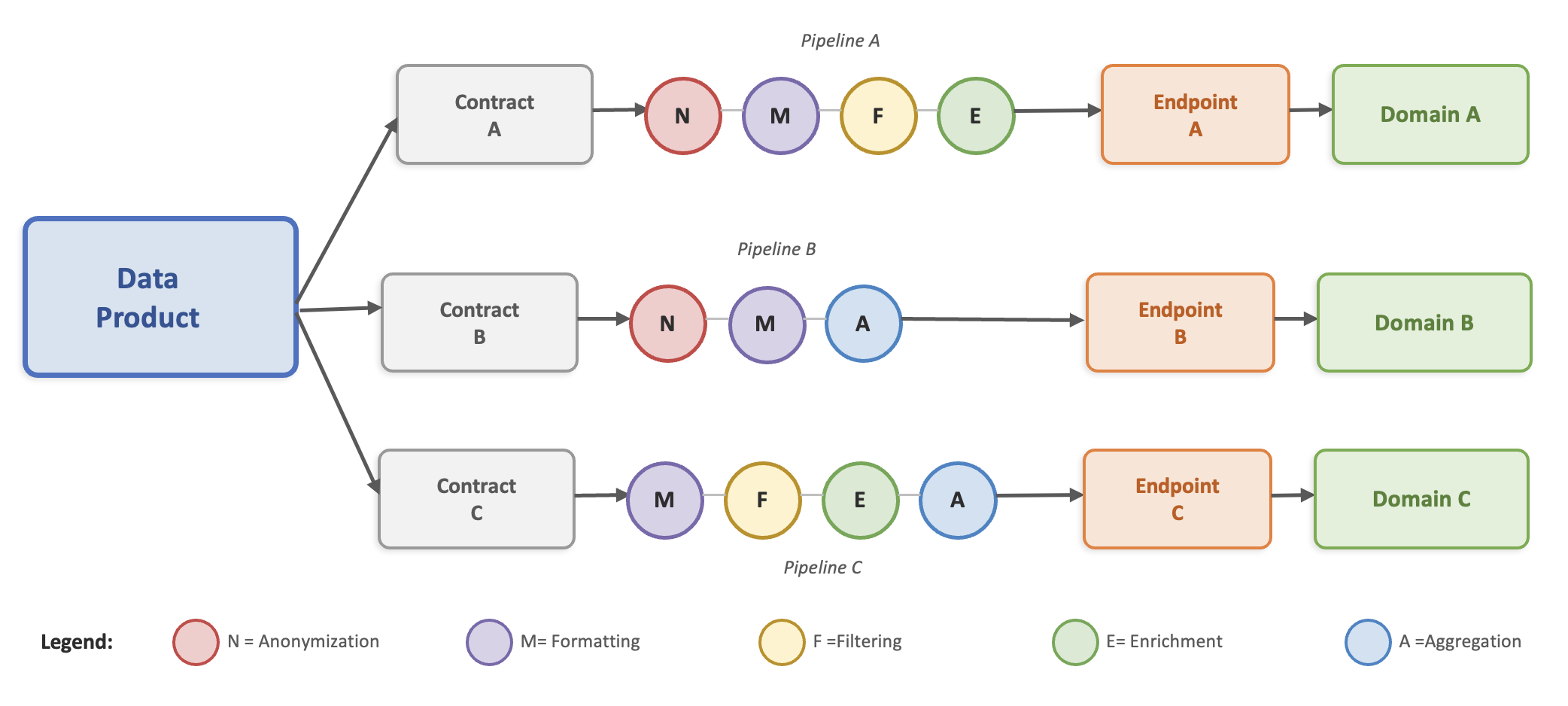}
\caption{A data product shared with consumers from different domains. Each data contract specifies transformation requirements, producing a dedicated pipeline. Stages are labeled by type, and similar coloring indicates similar transformation logic across pipelines.}
\label{fig:scenario}
\end{figure*}

Figure~\ref{fig:scenario} illustrates this intuition. A single data product is shared with three consumers, each governed by a distinct data contract and served by a dedicated pipeline. Despite their differences, these pipelines share substantial overlap: anonymization and formatting appear in both Pipeline~A and Pipeline~B, while formatting, filtering, and enrichment are common to Pipeline~A and Pipeline~C.  

In this paper, we investigate how such overlaps can be systematically exploited through reuse. We make the following contributions:

\begin{itemize}[leftmargin=*]
   \item We identify challenges and characterize reuse opportunities in federated data-sharing environments.
    \item We propose a reuse-oriented reference architecture for federated data-sharing environments.
   \item We present a preliminary evaluation of our reuse argument in a federated data-sharing scenario.
\end{itemize}

The remainder of this paper is organized as follows: First, we identify challenges and reuse opportunities in~\cref{sec:challenges} and outline a reuse-oriented reference architecture to address them in~\cref{sec:architecture}. We then partially implement this architecture and present a preliminary evaluation of our reuse argument in~\cref{sec:eval}. Lastly, we discuss related work in~\cref{sec:rw} before concluding in~\cref{sec:clf}.

\section{Problem Statement}\label{sec:challenges}\label{sec:reuse}Based on experiences and observations from pilot use cases in the domains of mobility, agriculture, and Industry~4.0 within the context of an EU-funded project~\cite{werner_teadal_2023}, as well as insights from state-of-the-art and survey studies~\cite{biswas2022art,epperson2022strategies,shojaee2024data}, we identified several challenges related to data-sharing pipelines in federated environments. In the following, we explain these challenges and categorize them into design-time and runtime challenges, arguing that addressing them requires systematic reuse across the data-sharing pipeline lifecycle.

\subsection{Challenges}
\vskip\baselineskip
\noindent \textit{Design-time Challenges:}

\textbf{Scalable Pipeline Construction (C1):} As the number of  consumers increases, providers must construct a growing number of data-sharing pipelines, each tailored to specific requirements. These pipelines are often created manually by selecting the transformation stages and the source data products~\cite{munappy2020,shojaee2024data}. In the absence of systematic support, this process does not scale: domain teams must repeatedly design similar pipelines from scratch, leading to redundant effort and inconsistent implementations. This highlights the need for mechanisms that enable systematic reuse of existing pipeline components and design~\cite{schultze2021data,dehghani2022}.

\textbf{Pipeline Evolution and Discoverability (C2):} 
Federated data sharing environments are inherently dynamic: new consumers join, requirements change, and new data products become available. As a result, pipelines cannot be treated as static artifacts. However, existing transformation logic and pipeline components are often difficult to discover and reuse, leading teams to re-implement functionality that already exists~\cite{epperson2022strategies}. Supporting pipeline evolution therefore requires visibility into available transformations and data products, as well as mechanisms to adapt and recombine them across pipelines~\cite{falconi2024,dehghani2022}.

\vskip\baselineskip
\noindent \textit{Runtime Challenges:}

\textbf{Redundant Execution (C3):}  Even when pipelines share identical or equivalent transformation logic, they are typically deployed and executed independently~\cite{epperson2022strategies,sedlak2024,falconi2024}. Without mechanisms to detect and exploit overlap at runtime, the platform 
may execute the same transformations multiple times on the same data. 
This leads to redundant computation, increased infrastructure costs, and unnecessary resource consumption.
Addressing this challenge requires runtime support for sharing and reusing transformation results and components across pipelines.

\textbf{System-level Trade-Offs (C4):} Design decisions made during pipeline construction, such as the choice of transformations or data sources, affect non-functional properties including latency, cost, and energy consumption.
These effects often only become visible at runtime and may interact across pipelines in non-obvious ways. As pipelines proliferate, managing such trade-offs becomes increasingly complex. Platforms therefore require mechanisms to observe and reason about the runtime behavior of pipelines and to optimize shared execution where possible~\cite{werner_teadal_2023}.

\textbf{Trust and Verifiability (C5):} In a federated environment, each transformation stage can be implemented and operated by different parties, making correctness, security, and compliance critical concerns. As the number of pipelines grows, verifying each implementation becomes increasingly difficult. This undermines trust in shared data products, particularly when sensitive data is involved~\cite{brito_trustops_2025}. Ensuring trustworthiness therefore requires verifiable evidence that transformations conform to their specified data contracts. Reusing well-defined and validated components can help reduce verification efforts and improve reliability~\cite{advocate,castillo_tcus_2025,wider2023}.

\subsection{The Case for Reuse}
The challenges identified above share a common root: federated data-sharing platforms, as introduced in~\cite{wider2025data,werner_teadal_2023}, lack mechanisms to recognize and exploit structural overlap across pipelines. As a result, similar transformation logic is repeatedly implemented and executed in isolation~\cite{biswas2022art,epperson2022strategies,pipeline-tool}. We argue that reuse — at both design-time and runtime — provides a unifying approach to address this limitation, aligning with established principles such as the FAIR~\cite{wilkinson2016} and data mesh guidelines~\cite{dehghani2022}, which emphasize standardized and reusable data assets. 

We therefore formulate the following hypothesis:

\begin{tcolorbox}[
    colframe=tealcolor,
    colback=white,
    boxrule=1.5pt,
    arc=4pt,
    left=6pt, right=6pt, top=6pt, bottom=6pt
]
\vspace{1em}
\textbf{Hypothesis} Enabling systematic \emph{reuse} both at design and runtime can address key challenges in federated data-sharing platforms.
\vspace{1em}
\end{tcolorbox}

\vskip\baselineskip
\noindent \textbf{Design-time Reuse}

At design time, reuse focuses on reducing the effort required to construct and evolve pipelines.

First, data-sharing pipelines often follow recurring structural patterns. Instead of designing pipelines from scratch, domain teams can instantiate and adapt reusable pipeline templates through patterns, reducing the design burden and improving consistency (C1).

Second, transformation logic can be encapsulated in reusable, parameterizable components. 
By maintaining a shared catalogue of available transformations, domain teams can discover and reuse existing logic when constructing or modifying pipelines, rather than implementing it. 
This improves scalability and supports pipeline evolution in dynamic environments (C2).

\vskip\baselineskip
\noindent \textbf{Runtime Reuse}

At runtime, reuse focuses on avoiding redundant computation and improving system-level efficiency.

When multiple pipelines require overlapping transformations, intermediate results can be materialized as shared data products and reused across consumers. 
By maintaining awareness of existing data products and deployed transformation services, the platform can route requests to already computed results where appropriate, reducing redundant execution and operational cost (C3). 

Such shared execution also enables better management of system-level trade-offs. By consolidating computation, platforms can optimize latency, cost, and resource utilization across pipelines, making these trade-offs more visible and controllable (C4).

Finally, reuse contributes to trustworthiness. Reusing well-defined and previously validated transformation components
reduces the need for repeated verification and the risk of errors. Trust can thus shift from individual pipelines to reusable components, improving reliability and compliance in federated environments (C5).

Together, design-time and runtime reuse form a coherent strategy for managing the complexity of federated data sharing. In the following section, we outline a reuse-oriented  architecture that operationalizes these principles.
\section{Reuse-Oriented Architecture}\label{sec:architecture}\begin{figure*}[t]
\centering
\includegraphics[width=0.6\textwidth]{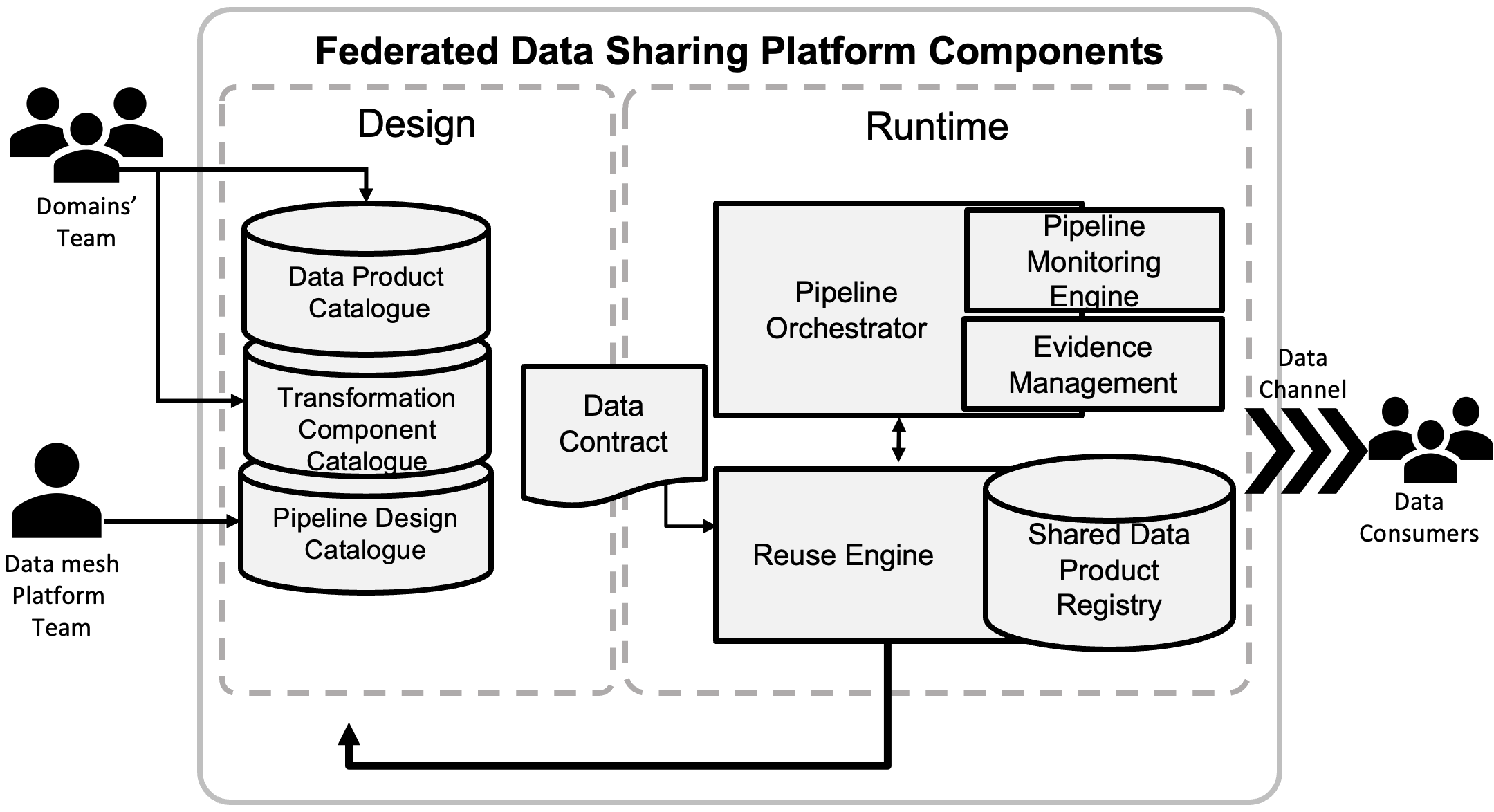}
\caption{Reuse-oriented architecture for federated data sharing environments.}
\label{fig:architecture}
\end{figure*}

This section outlines a reuse-oriented architecture for federated data-sharing environments. 
We specifically focus on enabling reuse both at design-time, by reusing transformation components, and at runtime, by sharing the intermediate data products~\cite{goedegebuure2024data} across different pipelines.
Figure~\ref{fig:architecture} depicts the proposed architecture of the federated data-sharing platform. 

In data mesh architecture, the data-sharing process between provider and consumer domains begins with defining the applicable policies, requirements and selecting the requested data product.
The outcome of this negotiation is formalized through a data contract, which specifies the requirements, schemas, and associated service-level agreements (SLAs)~\cite{wider2025data,jeffar2023federated}.

At this stage, domain teams~\cite{driessen2025} use the federated data-sharing platform to design, configure, and execute data-sharing pipelines that satisfy the agreements defined in the data contract.
The platform is part of the self-service data platform~\cite{dehghani2022,goedegebuure2024data}, providing the framework and infrastructure required by domains participating in the data mesh.

In the following, we present each major component separately in more detail, with a particular focus on its role in enabling reuse at both design time and runtime within the federated data-sharing platform.

\subsection{Enabling Reuse at Design-time}
The pipeline design catalogue and transformation component catalogue can be used to identify reuse during design time. 

The \emph{Data Product Catalogue} holds all available data products in the federation, and thus is the starting point for consumers to discover data products. 
The catalogue also holds the means to begin negotiation with the data provider to create a custom data sharing pipeline to expose the data to the consumer, e.g., adding required GDPR-compliant anonymization. 
It also holds all relevant information to determine the usage rights and access rules to the data products.
The Data Product Catalogue is continuously updated by domain teams~\cite{driessen2025} through the submission of metadata and addresses for new data products.

The \emph{Transformation Component Catalogue} is the central component for enabling reuse. 
The domain development teams are responsible for maintaining a catalogue of all available transformation components within the domain, including their functional descriptions, input and output schemas, configuration parameters, and metadata, and for publishing this information to the federated data-sharing platform (C2). Each registered component is independently deployable, e.g., implemented as a service deployable through a container image. The registry supports discovery queries, allowing domain teams or automated tools to find components that match specific transformation requirements, e.g., user anonymization.
When new components are contributed by domain teams, the registry captures their metadata in accordance with the federation's metadata model. Components are versioned, enabling the registry to track changes and ensure backward compatibility when stages are updated.
The domain teams~\cite{driessen2025} are responsible for maintaining the Transformation Component Catalogue.

The \emph{Pipeline Design Catalogue} provides a repository of pipeline cloud design patterns and templates that can be applied during pipeline construction. 
The orchestrator can select and apply these patterns non-intrusively, e.g., caching, without modifying the source code of transformation components~\cite{masoudi2026nonintrusive}, enabling pattern reuse across different pipeline architectures (C1). This non-intrusive application can, for example, be achieved through sidecar containers and service mesh configurations that wrap transformations with additional functionality at deployment time and dynamically update only the configuration of the transformation component that the pattern should apply between or into them.
The data mesh platform team~\cite{driessen2025} is responsible for maintaining the Pipeline Design Catalogue.

\subsection{Enabling Reuse at Runtime}
The \emph{Pipeline Orchestrator} executes the transformation stages. The required transformation stages, their execution order, and their connections are specified in the data contract. 
The orchestrator queries the Reuse Engine to identify existing data products and data pipelines that can be shared rather than re-deployed. When a required transformation is already running as part of another pipeline, the orchestrator can reference the shared instance, routing data through existing infrastructure rather than provisioning new resources (C1, C3). The orchestrator supports the non-intrusive application of pipeline design patterns without the need to change pipeline components.

The pipeline orchestrator integrates a \emph{Pipeline Monitoring Engine} to provide energy, usage, and performance metrics at the transformation stage level. These metrics support the Reuse Engine's impact assessment and enable domain teams to evaluate the consequences of their design decisions (C4).
The storage and profiling of these non-functional metrics can also benefit from being stored as tamper-proof evidence, for example using blockchain technologies, to make them verifiable and trustworthy in cases of conflicts regarding the defined SLAs in the data contracts.

Beyond runtime metrics, the monitoring component also captures, authenticates, and stores execution evidence in a tamper-resistant manner. \emph{Evidence management} supports the auditing process and enables trustworthy cross-organizational data sharing by allowing consumers and providers to independently verify that pipelines were executed and execution performance as agreed in their data contracts (C5).
Moreover, since a reused transformation propagates across multiple consumers, contracts, and organizations, any hidden defect, version drift, or misconfiguration scales proportionally with adoption. This makes determinism and verifiability essential properties, as they provide reproducible behavior, auditable evidence of execution context (version, configuration, and runtime environment), and accountability in compliance checks or disputes (C5) and also builds a common ground across the federation to share and act on observed runtime data.

The \emph{Reuse Engine} is the core decision-making component that identifies reuse opportunities and recommends reuse strategies at runtime.
It operates by analyzing the structural and semantic similarities of transformations across existing pipelines (C2, C3). At its core, the engine performs three steps. Firstly, given the set of currently deployed pipelines and a new data contract, the engine identifies which transformation components in the new pipeline \emph{overlap} with components in existing pipelines. This detection considers both the functional logic of stages and their configuration parameters (C3).
Secondly, the engine leverages the metrics collected by the pipeline monitoring engine to \emph{assess the impact} of potential decisions. This also includes observing implemented reuse to drive future recommendations. Domain teams can review this data to make informed decisions about when and how to apply reuse in the future and further tune recommendations.
However, based on detected overlaps, assessed impact, available transformation components, and shared data products, the reuse engine can recommend promoting overlapping pipelines to shared data products to avoid re-execution. Such shared data products are then registered in the \emph{Shared Data Product Registry} for reuse in future data-sharing pipelines.
Alternatively, the engine can identify frequently used transformation components and recommend deploying them as persistent, shareable platform services that can then be used across several otherwise non-overlapping pipelines, thereby avoiding potential cold-start issues. Such frequently used components are then also promoted in the transformation component catalogue.
These recommendations can be automatically applied by the pipeline orchestrator or first approved by domain teams before being enacted.
\section{Evaluation}\label{sec:eval}We conducted a preliminary study to assess the plausibility of the reuse hypothesis, targeting two core reuse mechanisms of the outlined architecture.
As a proof of concept, we selected an existing scientific data-sharing pipeline scenario and adapted it to incorporate reusable transformation components and shared data product reuse. 
We then examined whether the anticipated benefits are realizable in practice. The evaluation focuses on pipeline duration, throughput and resource utilization as system-level quality indicators, as these offer the most directly observable signal at this stage. Other anticipated benefits, such as trustworthiness, consistency, and reduced developer effort are left for future validation.
The scripts for the experimental treatments, along with the measured metrics and generated plots, are publicly available on GitHub\footnote{https://github.com/ISE-Research-Papers/Reuse-for-Data-Sharing-Pipelines}. 

\subsection{Experimental Protocol}
\begin{figure*}[t]
\centering
\includegraphics[width=0.6\textwidth]{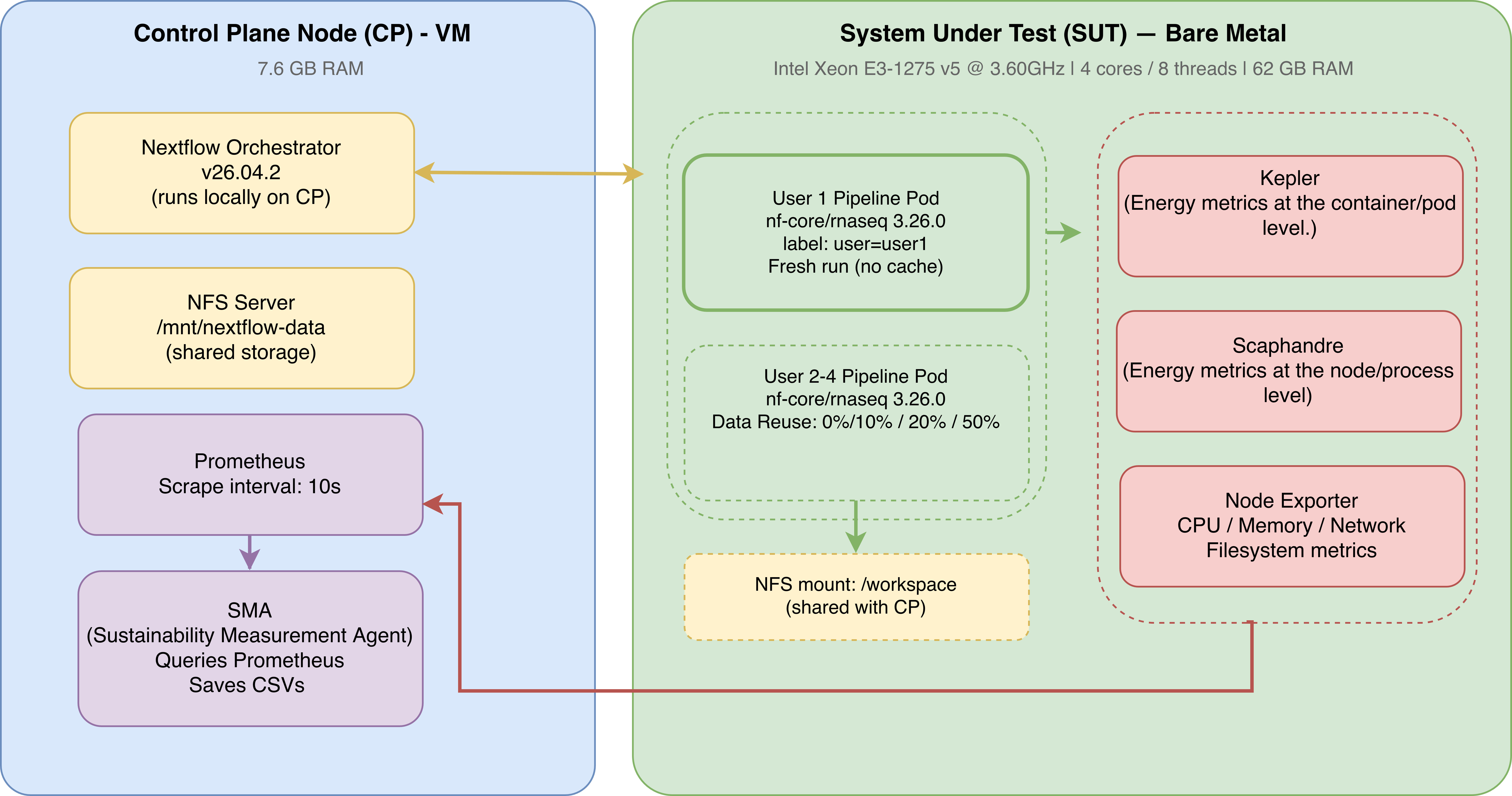}
\caption{Experiment setup infrastructure.}
\label{fig:experiment-setup}
\end{figure*}

\paragraph{Environment}
As illustrated in \Cref{fig:experiment-setup}, the experiment infrastructure consists of two nodes: a virtual machine acting as the control plane (CP) with 7.6 GB RAM, running the Nextflow orchestrator, NFS server, and Prometheus; and a bare-metal SUT node with an Intel Xeon E3-1275 v5 @ 3.60GHz (4 cores / 8 threads) and 62 GB RAM, where all pipeline pods execute under Kubernetes. Energy and resource metrics are collected on the SUT only via Kepler (per-container energy), Scaphandre (node-level RAPL power), and node-exporter (CPU/memory/network), all scraped by Prometheus and queried by the Sustainability Measurement Agent (SMA)~\cite{werner2025comprehensive} at the end of each experiment.

\paragraph{Workload} 
The scenario is based on the \texttt{nf-core/rnaseq} pipeline\footnote{https://nf-co.re/rnaseq/3.23.0/}, a community-maintained bioinformatics pipeline for RNA analysis that processes raw sequencing data through a number of transformation steps. 
We consider independent consumers each submitting analysis requests over the same twelve samples, with partially overlapping pipeline compositions.

\begin{figure*}[t]
\centering
\begin{subfigure}[t]{0.48\textwidth}
    \centering
    \includegraphics[width=\textwidth]{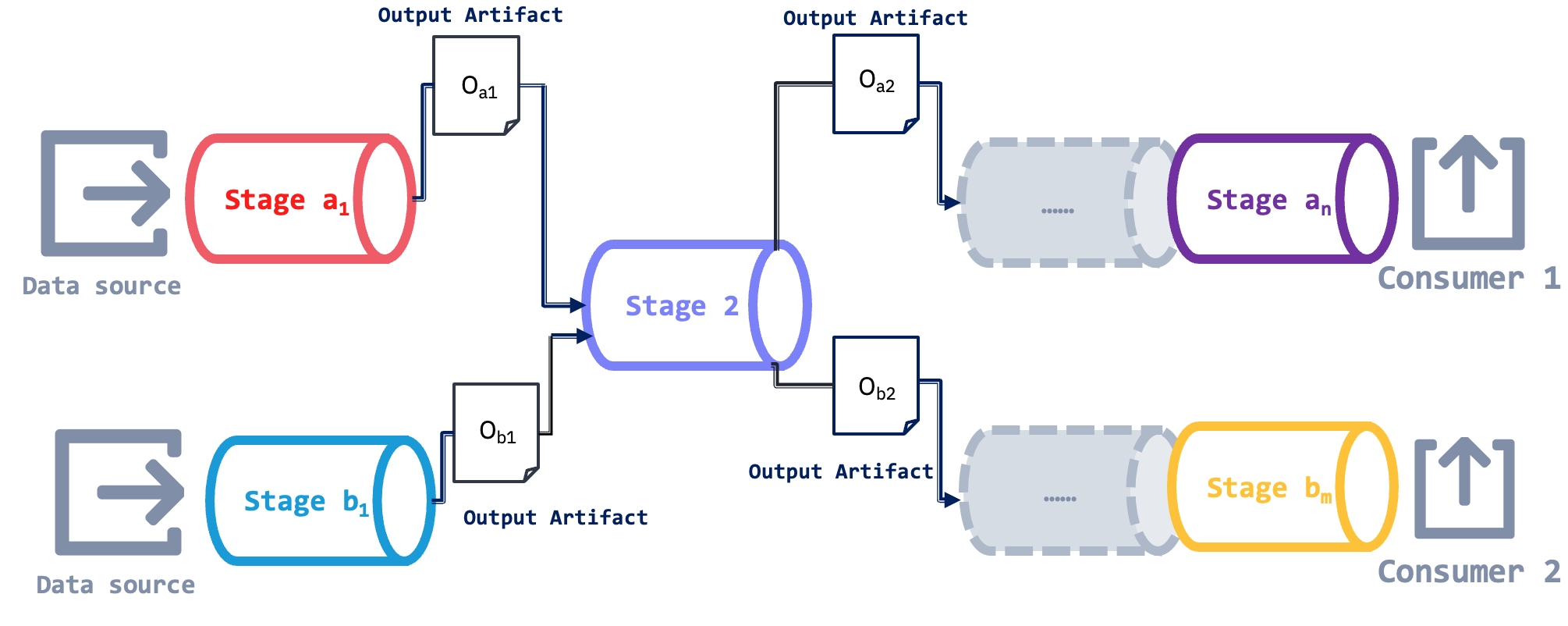}
    \caption{S2. Both consumers execute their pipelines independently; however, two transformation components run as long-lived Kubernetes pods and are not redeployed separately for each consumer.}
    \label{fig:arch2}
\end{subfigure}
\hfill
\begin{subfigure}[t]{0.48\textwidth}
    \centering
    \includegraphics[width=\textwidth]{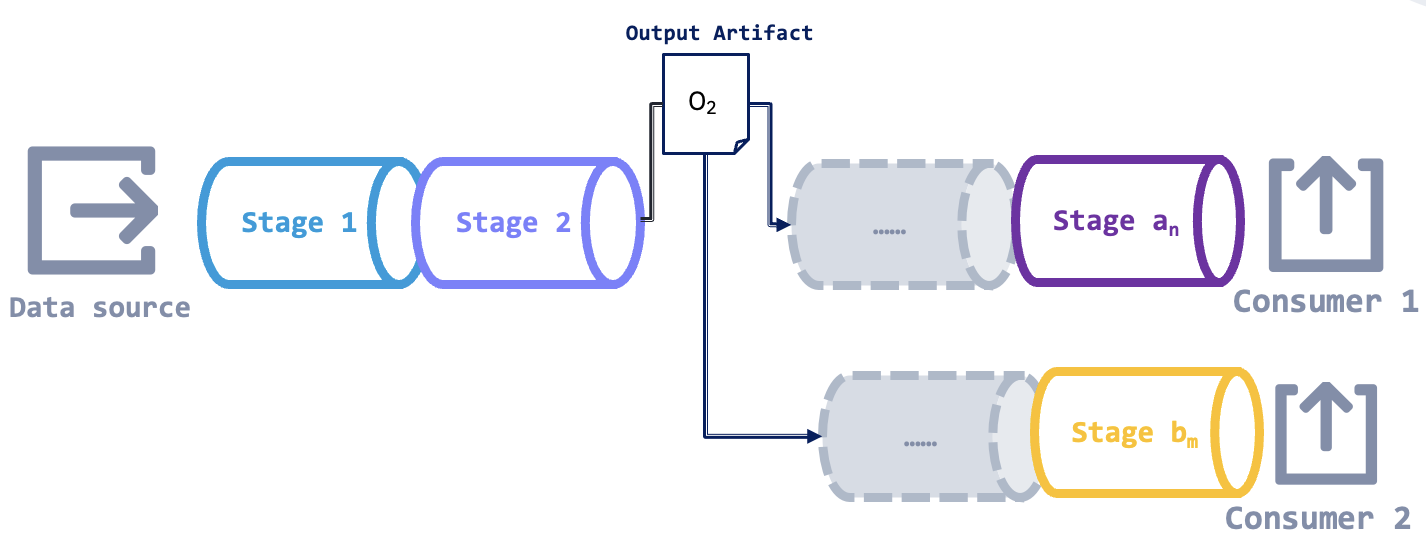}
    \caption{S3. Both consumers use intermediate results as a shared data product rather than re-executing those transformations.}
    \label{fig:arch1}
\end{subfigure}
\caption{Simplified demonstration of the experiment scenarios: S2 and S3.}
\label{fig:exp-scenario}
\end{figure*}

\paragraph{Treatment} 
We define the following three scenarios as treatments. \Cref{fig:exp-scenario} illustrates a simplified version of the treatment scenarios. Each of these scenarios was tested with two and four consumers.

\begin{itemize}
    \item \textbf{Baseline (S1):} The complete 12 sample pipeline was executed independently and sequentially from scratch for each consumer. No results, transformations, or containers were shared, in order to simulate a federated data-sharing setting in which reuse is not enabled. 
    
    \item \textbf{Design-time Reuse (S2):} The two transformations components (\texttt{featurecounts-service} and \texttt{stringtie-service}) run as persistent platform services and are shared across pipelines. The engine invokes these services instead of spawning new containers for the featureCounts and StringTie analysis stages, thereby eliminating container startup overhead.

    \item \textbf{Runtime Reuse (S3):} all consumers share pre-computed intermediate data products~\cite{guber2024privacy} from a prior pipeline execution. The first consumer runs the full pipeline fresh, while subsequent consumers reuse a partially of data representing 10\%, 20\%, or 50\% of the total data product(data samples), reducing redundant computation proportionally.
\end{itemize}

For the total number of the 548 pods per consumer (per pipeline), in the S3 scenario, we skip 55 pods for 10\%, 99 pods for 20\% and 274 pods for 50\% per user who uses the intermediate data product. 
In the S2 scenario, featureCounts and StringTie are chosen for transformation component reuse because they both suffer from significant container startup overhead of 2 to 5 minutes in Nextflow. 
Reusing these two transformation steps across pipelines means about 5\% of reuse per pipeline. 

\paragraph{Measurements}
As indicators of platform-level benefits, we examine three system-level metrics: pipeline duration, throughput and resource utilization measured as energy consumption. 
Pipeline duration reflects the end-to-end latency experienced by consumers, throughput captures the number of transformations the platform can process within a given time window, while energy consumption serves as a proxy for the computational cost incurred by the platform. 
All metrics are directly affected by redundant execution and are therefore well-suited to assess whether reuse leads to measurable efficiency gains.
We utilized Prometheus and Kepler to collect these metrics. 
All resource consumption values are calculated over the interval between the first and last Nextflow activity on the SUT node, excluding the resource consumption associated with Nextflow orchestration and Kubernetes control plane activities executed on the CP node. 

\subsection{Results}
In the following, we present the results of our preliminary evaluation separated by the type of reuse.

\subsubsection{Design-time Reuse (S2)}
The design-time reuse (S2) evaluates the impact of reusing transformation components by replacing short-lived, per-transformation pipeline pods with long-running, shared service pods.
As mentioned, we deployed the featureCounts and StringTie transformation components as long-running pods, while all other pipeline steps are executed as standard Kubernetes Jobs.
A separate pipeline is maintained for each user sequentially.

\paragraph{Energy Consumption}

\begin{figure}[htbp]
\centering
\includegraphics[width=0.8\columnwidth,trim={0 2em 0 0},clip]{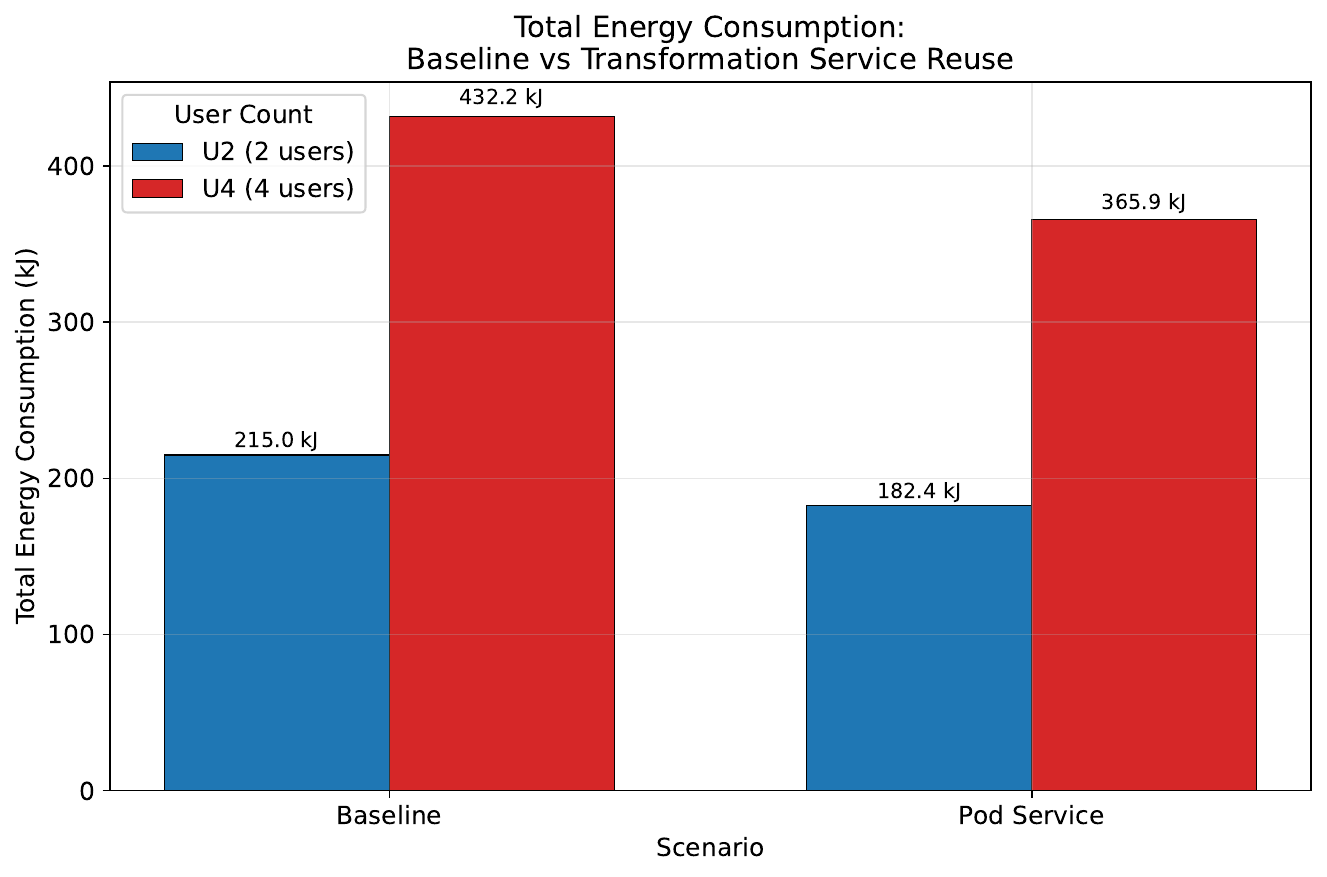}
\caption{Total node energy consumption (kJ) for baseline and transformation component reuse scenarios across 2 and 4 user counts.}
\label{fig:S2-energy_comparison}
\end{figure}

\Cref{fig:S2-energy_comparison} presents total node energy consumption across baseline and transformation component reuse scenarios. The baseline consumed 214.99,kJ (U2) and 432.18,kJ (U4). 
With transformation component reuse under sequential execution, energy drops to 182.41,kJ for U2 and 365.87,kJ for U4, representing savings of 15.2\% and 15.3\% respectively.

\paragraph{Pipeline Duration}

\begin{figure}[htbp]
\centering
\includegraphics[width=0.8\columnwidth,trim={0 2em 0 0},clip]{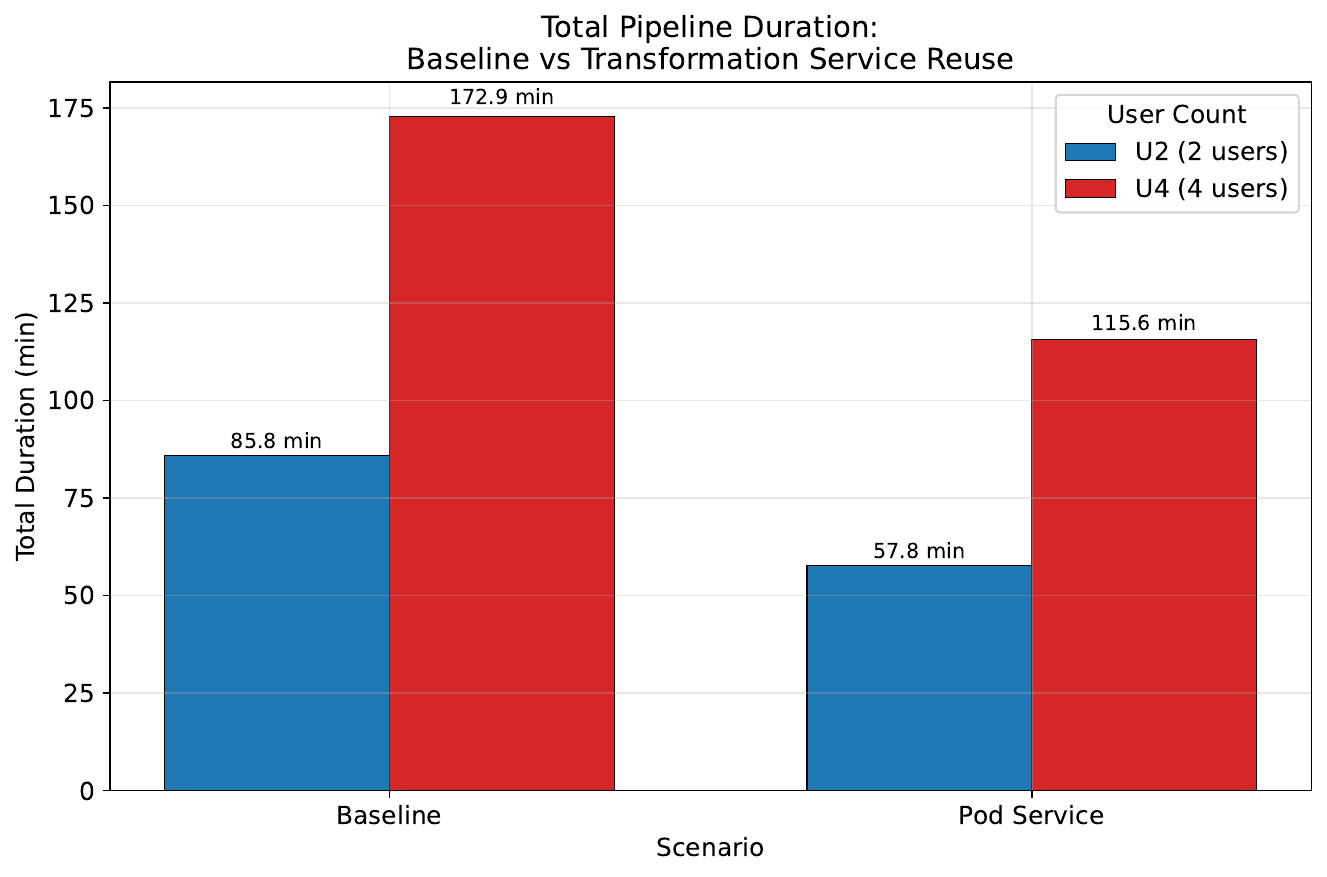}
\caption{Total pipeline duration (minutes) for baseline and transformation component reuse scenarios across 2 and 4 user counts.}
\label{fig:s2-duration_comparison}
\end{figure}

\Cref{fig:s2-duration_comparison} shows total pipeline duration for both scenarios. Baseline durations are 85.78,min (U2) and 172.89,min (U4). With service reuse, durations are reduced to 57.78,min and 115.60,min respectively, corresponding to savings of 32.6\% (U2) and 33.1\% (U4). 
The duration savings are substantially larger than the energy savings, indicating that pod startup and scheduling latency, rather than computation itself, accounts for a fraction of total pipeline time. 

\paragraph{System Throughput}

\begin{figure}[htbp]
\centering
\includegraphics[width=0.8\columnwidth,trim={0 2em 0 0},clip]{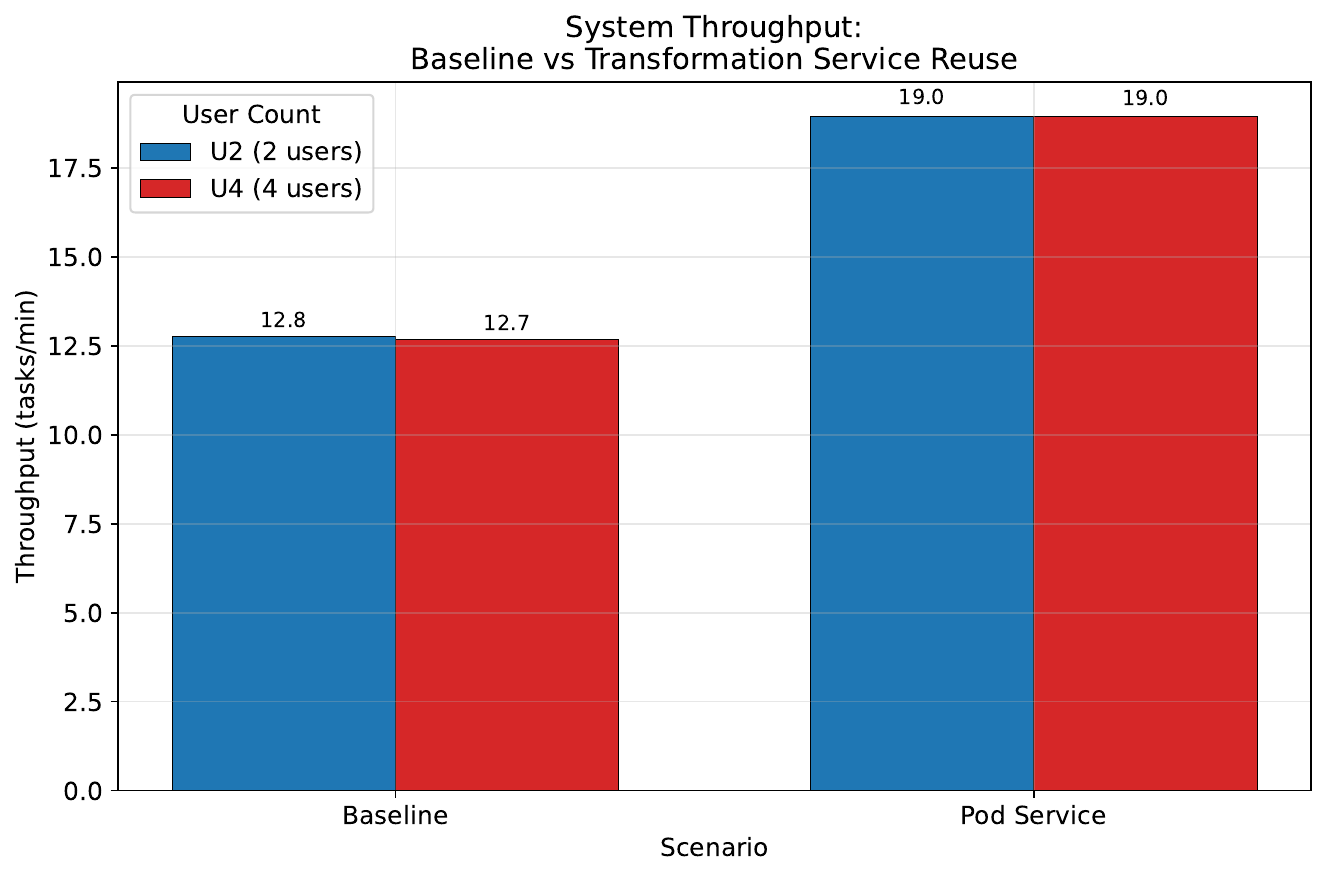}
\caption{System throughput (TC/min) for baseline and transformation components reuse scenarios across 2 and 4 user counts.}
\label{fig:s2-throughput_comparison}
\end{figure}

\Cref{fig:s2-throughput_comparison} presents system throughput measured in transformation components (TC) per minute. Baseline throughput is 12.78 TC/min (U2) and 12.68 TC/min (U4). With service reuse, throughput increases to 18.97 TC/min and 18.96 TC/min respectively, representing gains of 48.5\% (U2) and 49.6\% (U4). 

\subsection{Runtime Reuse (S3)}

The runtime reuse (S3) evaluates the effect of reusing pre-computed intermediate data products on total energy consumption, pipeline duration, and system throughput.
A separate pipeline is maintained for each user.

\begin{figure}[htbp]
\centering
\includegraphics[width=0.8\columnwidth,trim={0 2em 0 0},clip]{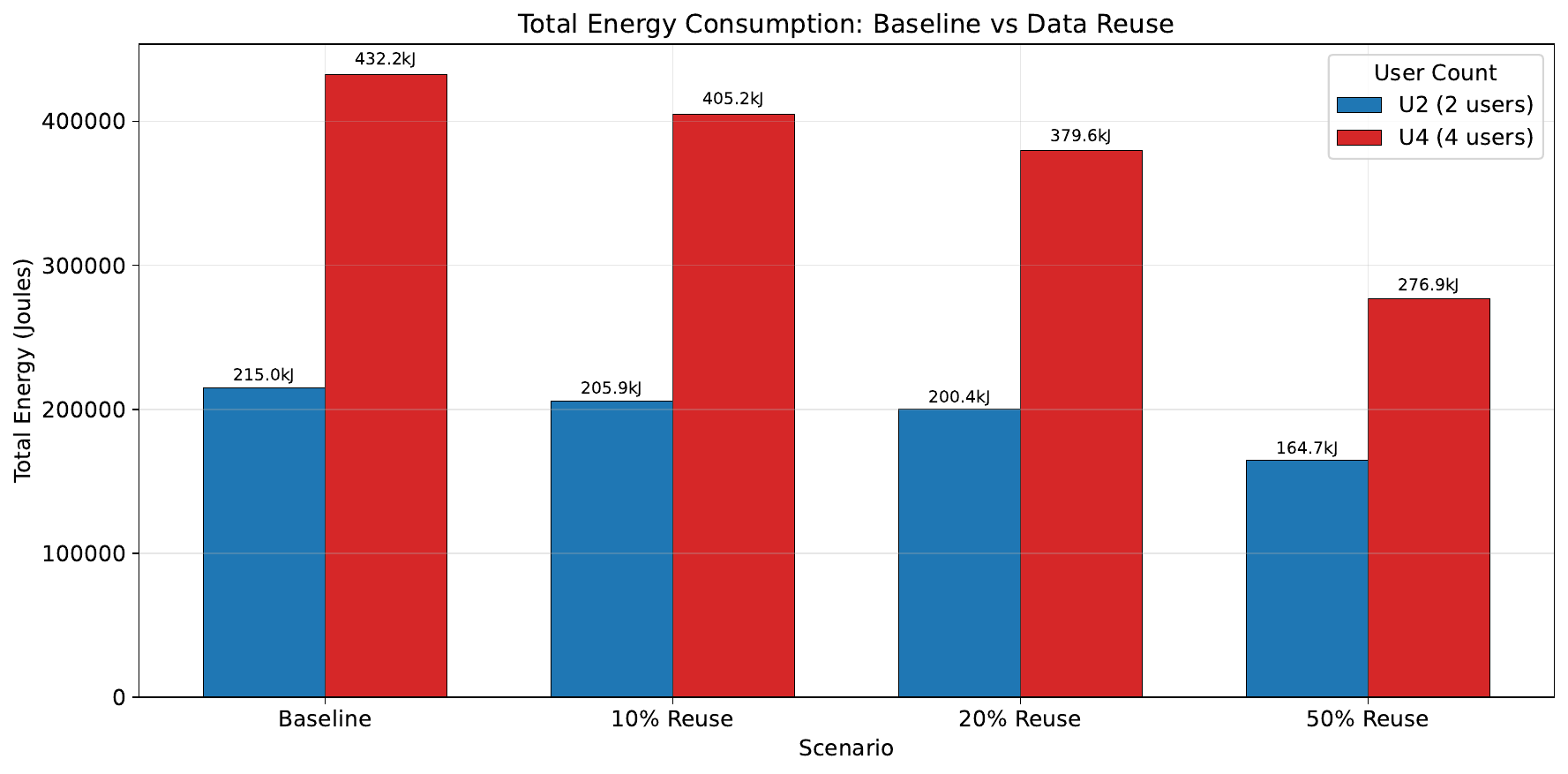}
\caption{Total node energy consumption (kJ) for baseline and data reuse scenarios across 2 and 4 user counts.}
\label{fig:S3-energy_comparison}
\end{figure}

\paragraph{Energy Consumption}
\Cref{fig:S3-energy_comparison} presents the total node energy consumption measured across all scenarios. 
The baseline experiments consumed 214.99 kJ (U2) and 432.18 kJ (U4), reflecting the full computational cost of executing the pipeline for all users without any reuse. 
As the reuse percentage increases, energy consumption decreases consistently for both user counts. For U2, energy consumption decreases by 4.2\% at 10\% reuse (205.88 kJ), 6.8\% at 20\% reuse (200.40 kJ), and 23.4\% at 50\% reuse (164.67 kJ). 
For U4, the savings are more pronounced, with reductions of 6.3\% at 10\% reuse (405.16 kJ), 12.2\% at 20\% reuse (379.59 kJ), and 35.9\% at 50\% reuse (276.94 kJ). 
These results demonstrate that energy savings scale with the reuse percentage, with 50\% reuse yielding larger savings than 10\% or 20\%.

\begin{figure}[htbp]
\centering
\includegraphics[width=0.8\columnwidth,trim={0 2em 0 0},clip]{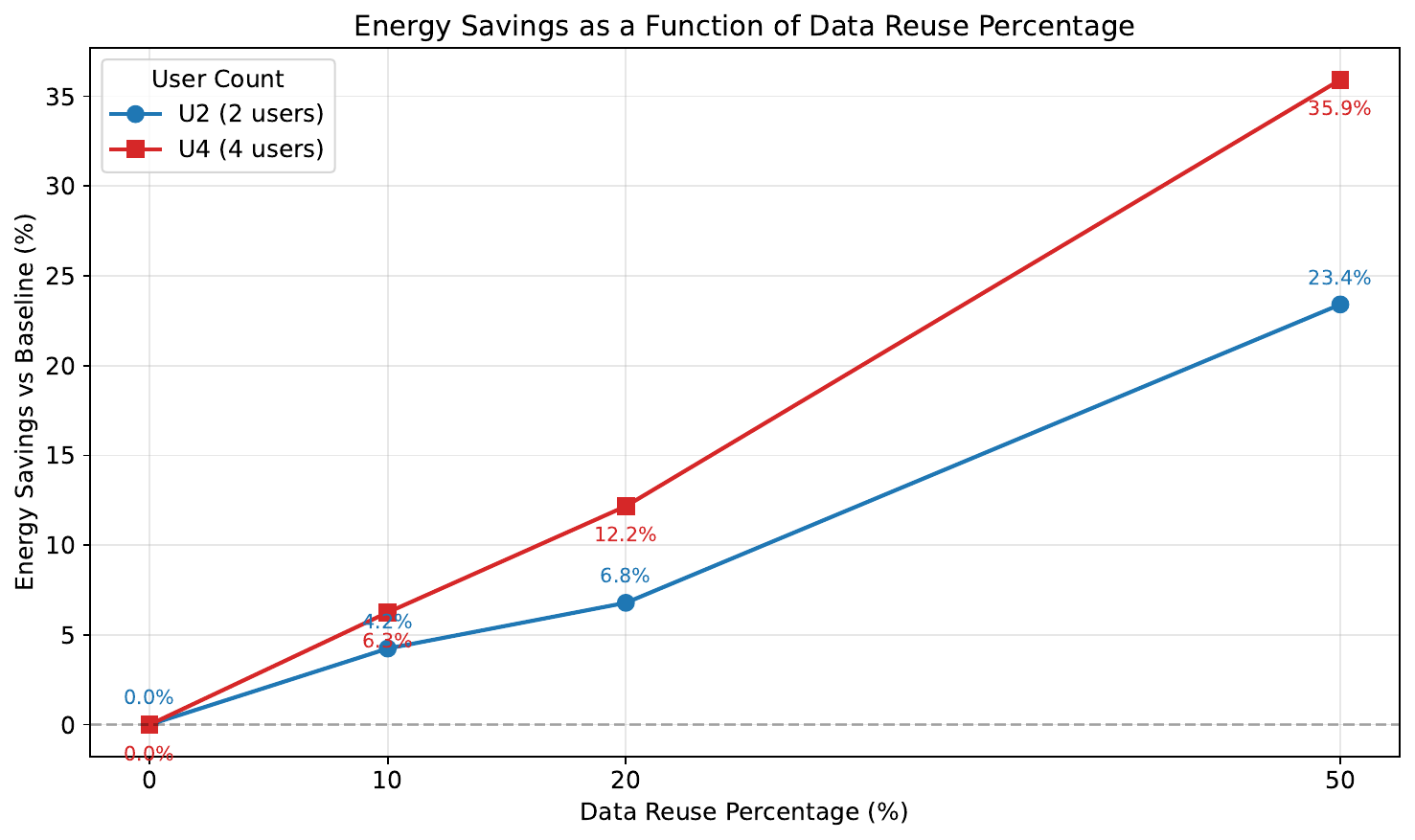}
\caption{Energy savings relative to the sequential baseline as a function of the reuse percentage, for 2 and 4 user counts.}
\label{fig:S3-energy_savings_vs_reuse}
\end{figure}

\Cref{fig:S3-energy_savings_vs_reuse} illustrates the energy savings as a function of the reuse percentage for both user counts. 
The U4 configuration consistently achieves higher energy savings than U2 at equivalent reuse levels, suggesting that the benefit of reuse amplifies with the number of concurrent users sharing the intermediate data products.

\paragraph{Pipeline Duration}

\begin{figure}[htbp]
\centering
\includegraphics[width=0.8\columnwidth,trim={0 2em 0 0},clip]{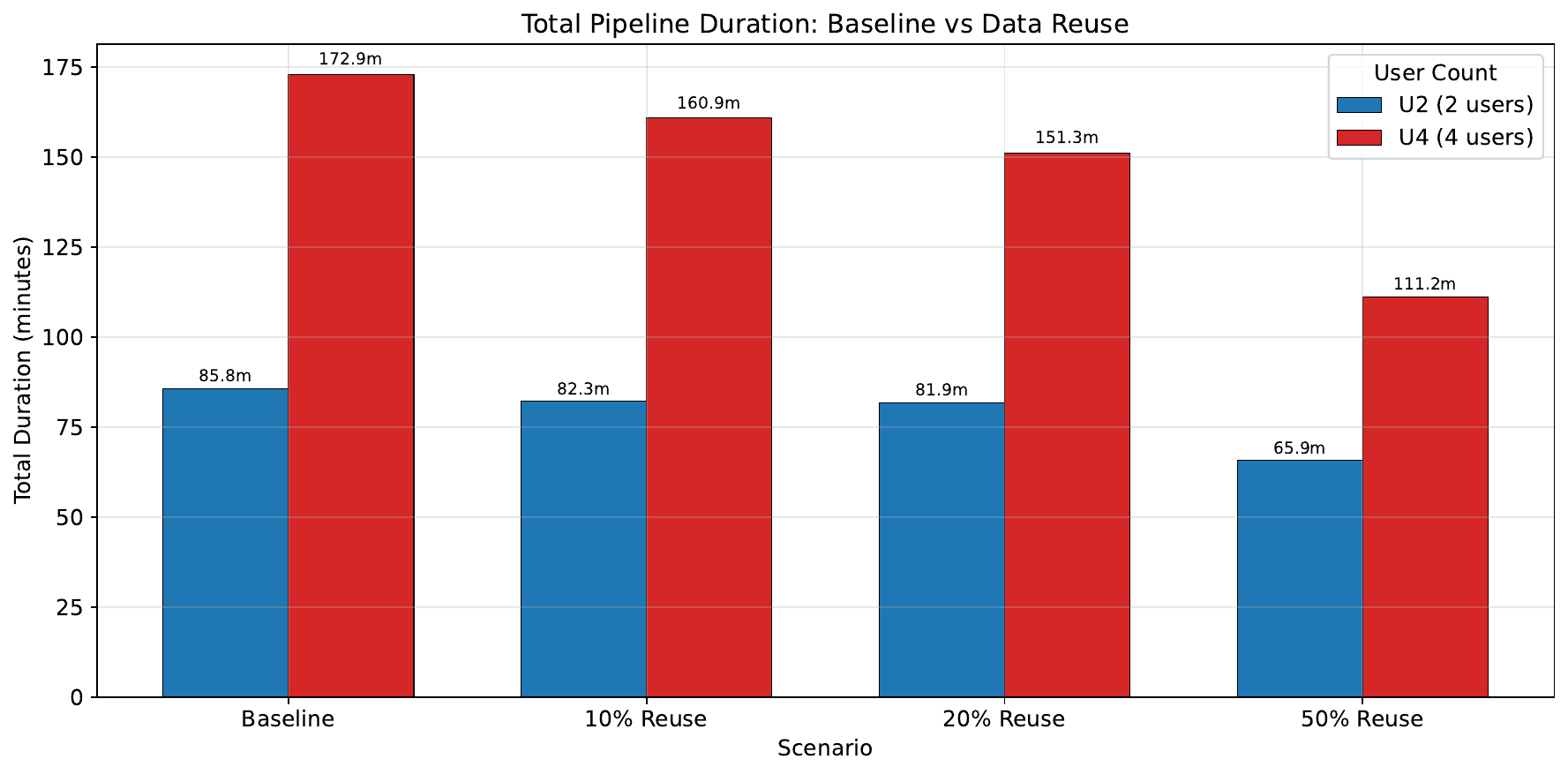}
\caption{Total pipeline duration (minutes) for baseline and data reuse scenarios across 2 and 4 user counts.}
\label{fig:duration_comparison}
\end{figure}

\Cref{fig:duration_comparison} shows the total pipeline duration across all scenarios. The baseline durations are 85.78,min (U2) and 172.89,min (U4). Reuse reduces duration in proportion to the percentage of skipped transformation components due to reuse: for U2, duration decreases by 4.1\%, 4.5\%, and 23.2\% at 10\%, 20\%, and 50\% reuse respectively. 
For U4, the corresponding reductions are 6.9\%, 12.5\%, and 35.7\%. The duration savings closely mirror the energy savings, confirming that the primary driver of both metrics is the reduction in active computation time rather than other factors such as I/O or scheduling overhead. 
At 50\% reuse with U4, the total pipeline duration drops from 172.89,min to 111.21,min, representing a saving of over one hour of wall-clock time.

\paragraph{System Throughput}

\begin{figure}[htbp]
\centering
\includegraphics[width=0.8\columnwidth,trim={0 2em 0 0},clip]{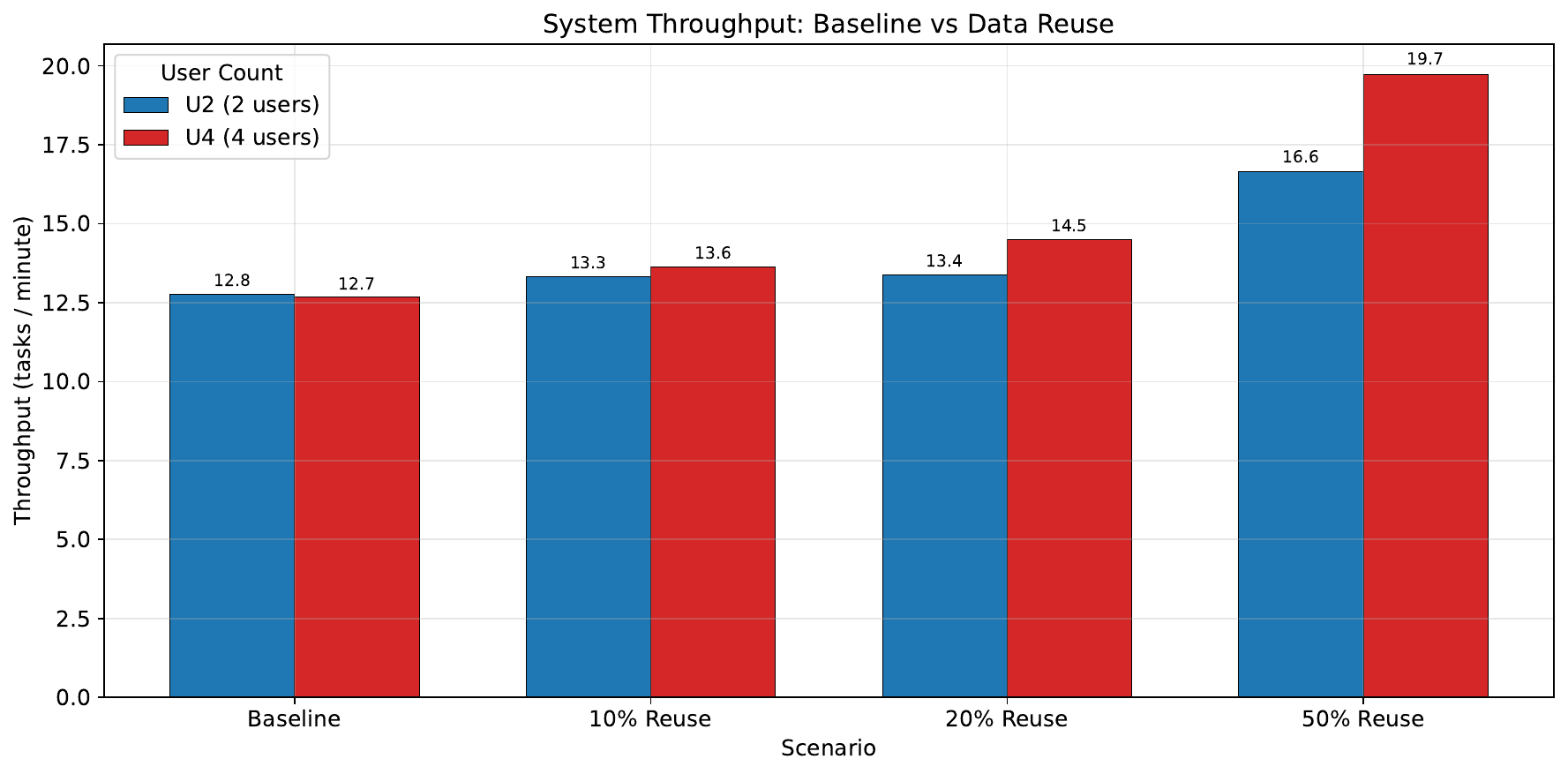}
\caption{System throughput (TC/min) for baseline and data reuse scenarios across 2 and 4 user counts. Each user executes 548 transformation component.}
\label{fig:throughput_comparison}
\end{figure}

\Cref{fig:throughput_comparison} presents system throughput, measured as the number of pipeline transformation components (TC) completed per minute across all users in each scenario. Baseline throughput is 12.78 TC/min (U2) and 12.68 TC/min (U4), reflecting near-identical per-user performance regardless of user count under sequential execution. 
As reuse percentage increases, throughput improves substantially. For U2, throughput increases by 4.2\%, 4.7\%, and 30.2\% at 10\%, 20\%, and 50\% reuse respectively. 
For U4, gains are 7.5\%, 14.3\%, and 55.5\%. The U4 50\% reuse scenario achieves the highest throughput of 19.71 TC/min, representing a 55.5\% improvement over the U4 baseline. 
This result highlights that intermediate data product reuse not only reduces energy consumption but also substantially increases the effective processing capacity of the system, particularly at higher reuse levels and larger user counts.

\subsection{Discussion}

The runtime reuse results (S3) confirm this in our setting. Energy consumption, pipeline duration, and throughput all improve monotonically with the reuse percentage, and the benefit amplifies with the number of concurrent users. 
This is expected, because runtime reuse works by skipping the recomputation of intermediate data products already materialized by another pipeline; the savings are therefore bounded by the fraction of overlapping transformation components, which is why duration savings closely track energy savings, as both stem from the same reduction in active computation. 
Crucially, this form of reuse is well supported by current orchestrator tooling, since persisting and resolving intermediate data products is a capability of existing dataflow engines.

However, runtime reuse is fundamentally constrained by the requirement that consumers share overlapping pipeline structure. A pre-computed product can only be reused if another pipeline requests the same transformation on the same upstream inputs under the same requirements. 

\begin{tcolorbox}[
    colframe=slateblue,
    colback=white,
    boxrule=1.5pt,
    arc=4pt,
    left=6pt, right=6pt, top=6pt, bottom=6pt
]
Reusing overlapping pipelines as persisted shareable data products has already shown promising results in existing dataflow engines at runtime. 
\end{tcolorbox}

Design-time reuse (S2) relaxes this constraint. By promoting transformation components to long-running shared service pods, the unit of reuse becomes the transformation component itself rather than a materialized data product. Two users therefore benefit from the same shared component as long as their pipelines overlap in that transformation component, regardless of whether the surrounding pipeline structure, upstream inputs, or requirements coincide. This is a substantially weaker and more frequently satisfied precondition than the full-prefix overlap required by runtime reuse, making design-time reuse applicable in heterogeneous settings where users share processing logic but not entire pipelines.

Design-time reuse is advantageous precisely when the orchestration overhead of a pod is large relative to its execution time. 
Runtime reuse, which eliminates the computation entirely, is the more effective lever. The two strategies are therefore complementary: design-time reuse targets orchestration overhead and tolerates pipeline heterogeneity, whereas runtime reuse targets redundant computation but requires pipeline congruence.

While current orchestrators expose data-product reuse as an option, there is no built-in option to declare a transformation as a long-lived shared service.
Design-time reuse must therefore be built and maintained by hand, outside the orchestrator's abstractions, including managing the lifecycle of the long-running components and, critically, the logic for scaling them up and down with demand, which the orchestrator already provides automatically for per-task pods but not for persistent services.

\begin{tcolorbox}[
    colframe=slateblue!30,
    colback=white,
    boxrule=1.5pt,
    arc=4pt,
    left=6pt, right=6pt, top=6pt, bottom=6pt
]
Implementing reusable transformation components is possible, but it currently requires costly and inefficient workarounds for design-time reuse.
\end{tcolorbox}

\subsection{Threats to Validity}

Overall, we evaluated the feasibility of \emph{reuse} in data pipelines, with promising results for platform-level efficiency in terms of energy consumption and pipeline performance.

Our evaluation considers reuse scenarios with only two and four consumer pipelines. While this allows us to study the effect of reuse under different system loads, the effectiveness of reuse may also depend on pipeline structure, dependencies, and resource usage patterns. Investigating how pipeline characteristics influence reuse effectiveness remains future work.

In addition, this work focuses on runtime metrics, particularly pipeline performance and energy consumption. Other aspects, such as the engineering effort required to design, implement, and maintain reuse mechanisms, were not evaluated. Similarly, potential benefits including improved trustworthiness, reliability, consistency, and reduced implementation effort remain to be studied.

Finally, our experiments were conducted in a controlled environment rather than a fully federated setting. Although we expect the observed reuse benefits to generalize to federated environments, this assumption still requires validation in real-world deployments.

\section{Related Work}\label{sec:rw}Data mesh was introduced by Dehghani~\cite{dehghani2022} and later consolidated in work that clarifies its principles, architectural elements, and open challenges~\cite{machado2022,goedegebuure2024data}. Beyond conceptual framing, Eichler et~al.~\cite{eichler2022data} discuss the provider-side shift from data assets to data products and highlight effort and incentive issues that affect adoption in practice. Governance is another central theme: Wider et~al.~\cite{wider2023} examine decentralized governance mechanisms in data mesh platforms, and later work explores AI-assisted support for governance and compliance decisions~\cite{wider2025}. For federated settings, Sedlak et~al.~\cite{sedlak2024} are particularly close to our context, combining data mesh and serverless exchange and defining a five-phase data product lifecycle. We build on that lifecycle view and focus specifically on reuse across pipeline phases.

Falconi et~al.~\cite{falconi2024} frame trustworthy data sharing through the notion of \emph{data friction}, i.e., technical obstacles that accumulate during exchange. Our work complements this perspective by treating reuse as an architectural mechanism to reduce friction at concrete pipeline stages. This positioning is also informed by software reuse literature: reuse has long been studied in software engineering~\cite{caldiera2002}, including quantitative productivity benefits~\cite{gaffney1989} and privacy-aware reuse constraints~\cite{guber2024privacy}. However, most prior work emphasizes development-time gains, while we examine reuse as a lever for runtime-relevant qualities in federated data-sharing pipelines, such as resource consumption and operational efficiency. In parallel, the FAIR principles~\cite{wilkinson2016} established reusability as a cornerstone of data stewardship; our contribution extends that perspective from datasets to the pipeline infrastructure that transforms, orchestrates, and delivers data products.

The contributions presented in this paper build on a sequence of our prior publications that addressed complementary aspects of reuse in data-sharing pipelines. A first line of work focused on resource awareness. In~\cite{masoudi2025energy}, we introduced an energy profiling model for data-sharing pipelines and used it to derive initial reuse strategies based on estimated energy consumption. This established the basic premise that reuse in pipeline design is not only a matter of engineering convenience, but can also be motivated by measurable operational effects. We extended this perspective in~\cite{masoudi2025preshare} with \emph{Pre-share Data}, an assistance tool that supports domain teams during pipeline configuration by making resource-aware design decisions more explicit and actionable.

A second line of work addressed how reuse can be realized across heterogeneous technical settings. In~\cite{masoudi2026nonintrusive}, we proposed a non-intrusive framework for the deferred integration of cloud design patterns into data-sharing pipelines. That work showed how transformation stages and architectural solutions can be reused across different pipeline implementations without modifying service source code, thereby lowering the adoption barrier for reuse in practice.

A third line of work concerned trustworthiness, verifiability, and governance in pipeline-based data sharing. In~\cite{advocate}, we presented \emph{Advocate}, an agent-based system for generating attested evidence of cloud-native application operations. Related work explored how verifiable evidence can be produced during pipeline creation~\cite{brito_trustops_2025,castillo_tcus_2025} as well as during runtime operation~\cite{castillo_slas_2025}. We also investigated governance- and sovereignty-oriented pipeline designs~\cite{brito2025synthgard}, which are particularly relevant in federated settings where technical reuse must remain compatible with organizational control and accountability requirements.

Taken together, these prior contributions cover three concerns that are usually treated separately: resource-aware optimization, technical realization of reuse across heterogeneous infrastructures, and verifiable trustworthy operation. The present paper synthesizes these threads into a unified perspective. Rather than treating reuse as an isolated optimization technique, we frame it as an overarching design strategy for federated data-sharing pipelines and use the lessons from these earlier works to motivate the proposed reference architecture, including the points at which verifiable evidence and governance mechanisms need to be anchored.

\section{Conclusion and Future Work}\label{sec:clf}In this paper, we argued that \emph{reuse} should serve as a guiding principle for addressing the growing complexity of data-sharing pipelines in federated data sharing environments.

We identified five challenges that arise at both design time and runtime of data-sharing pipelines. On this basis, we proposed a reuse-oriented reference architecture, which enables domain teams to compose, share, and evolve pipelines from reusable components. 
Our preliminary evaluation assesses the applicability of reuse at both design time and runtime. 
We found that runtime reuse of shared data products reduces pipeline execution time and energy consumption, and can already be realized with existing pipeline orchestrators. In contrast, design-time reuse of transformation components is not adequately supported by current orchestrators, and implementing it through workarounds incurs adverse performance and energy overhead.

Future work will focus on integrating reuse more natively into execution engines and orchestration frameworks, reducing the need for external implementation and improving overall efficiency.

\section*{Acknowledgements}
\footnotesize \noindent 
Funded by the Federal Ministry of Research, Technology, and Space (BMFTR) as part of the GANGES2 project (16KIS2609). Views and opinions expressed are, however, those of the author(s) only and do not necessarily reflect those of the BMFTR. Neither the BMFTR nor the granting authority can be held responsible for them. Supported by the Estonian Centre of Excellence in AI (EXAI), funded by the
Estonian Ministry of Education and Research.

\bibliographystyle{IEEEtran}
\bibliography{ref_2}

\end{document}